\documentclass{article}
\usepackage{spconf,amsmath,graphicx}
\usepackage{xcolor}
\usepackage[font={small}]{caption}
\usepackage{cite}

\usepackage{ulem} 
\usepackage{fancyhdr}

\title{WaveNILM: A CAUSAL NEURAL NETWORK FOR POWER DISAGGREGATION FROM THE COMPLEX POWER SIGNAL}
\name{Alon Harell, Stephen Makonin, and Ivan V. Baji\'{c}\thanks{This work was funded in part by IC-IMPACTS.}}
\address{School of Engineering Science, Simon Fraser University, Burnaby, Canada \\
Email: aharell@sfu.ca, smakonin@sfu.ca, ibajic@ensc.sfu.ca}
\begin{document}
%\ninept
%
\maketitle

\thispagestyle{empty}
\renewcommand{\headrulewidth}{0.0pt}
\thispagestyle{fancy}
\lhead{}
\chead{Accepted at the 44th International Conference on Acoustics, Speech, and Signal Processing (ICASSP).}
\rhead{}
\lfoot{}
\cfoot{Copyright \copyright{ } 2019 IEEE. The original publication is available for download at ieeexplore.ieee.org.}
\rfoot{}

\begin{abstract}
Non-intrusive load monitoring (NILM) helps meet energy conservation goals by estimating individual appliance power usage from a single aggregate measurement.
Deep neural networks have become increasingly popular in attempting to solve NILM problems; however, many of them are not causal which is important for real-time application. We present a causal 1-D  convolutional neural network inspired by WaveNet for NILM on low-frequency data. We also study using various components of the complex power signal for NILM, and demonstrate that using all four components available in a popular NILM dataset (current, active power, reactive power, and apparent power) we achieve faster convergence and higher performance than state-of-the-art results for the same dataset. 
\end{abstract}
\begin{keywords}
NILM, power disaggregation, convolutional neural network, causality
\vspace{-0.2cm}
\end{keywords}

\ULforem

\section{Introduction}
\vspace{-0.2cm}
\label{sec:intro}

As the cost and environmental impact of energy use continues to increase, the importance of power conservation and planning is growing significantly. Consumers and providers alike are encouraged to reduce power usage by many factors such as rising costs, legislation, and public image. Non-intrusive load monitoring (NILM), first proposed by Hart in 1992~\cite{hart}, suggests disaggregating a single measurement of power from a central meter (e.g., a household smart meter) to estimate the  power usage of different loads. Specifically, the aggregate power measurement $s(t)$ is given by
\vspace{-0.2cm}
\begin{equation}
s(t) = \sum_{k=1}^{K}s_k(t) + n(t),    
\vspace{-0.1cm}
\end{equation}
where $s_k(t)$ is the power level of the $k$-th appliance and $n(t)$ is noise. The goal is to estimate $s_k(t),k=1,2,...,K,$ from $s(t)$, which is a form of \textit{source separation}. This information can be used by the consumer to make informed decisions and by providers to educate clients and better plan production. 

In the majority of NILM solutions, only one electrical measurement is considered, usually active/real power or current, and the output for each appliance is either a state classification (which is then related to a power level) or a direct regression of power used. We examine the effect of using multiple aggregate electric measurements (still collected by one central meter) to improve performance and convergence speed. 
\vspace{-0.3cm}
\subsection{Active and reactive power}
\label{ssec:power}

Power utility companies are generally concerned with two types of loads -- active power ($P$) loads and reactive power ($Q$)  loads. Active power loads (e.g., an electric stove) receive power from the grid and perform work dissipating it. Reactive power loads (e.g., a capacitor) receive power from the grid, store it, and after some period release it back to the grid in the opposite direction without dissipation. In many cases loads are both active and reactive -- such as a heat pump or an air conditioning unit. % which contain a spinning fan.

Mathematically, active power results from in-phase voltage and current, whereas reactive power results from out-of-phase voltage and current. %Power factor (PF) indicates how out-of-phase voltage and current are. $\text{PF} = 1$ indicates that voltage and current are completely in-phase. 
Apparent power $S$, sometimes referred to as total momentary power, can also be a useful cue for disaggregation. These quantities are related as:
\vspace{-0.1cm}
\begin{equation}
S = I\cdot V , \quad 
P = S \cdot \cos(\theta) , \quad
Q = S \cdot \sin(\theta) ,
\vspace{-0.1cm}
\end{equation}

%$$
%\text{PF} = \frac{P}{S} = \cos(\theta) ,
%$$
where $I$ is current, $V$ is voltage, and $\theta$  is the phase of voltage relative to current (i.e., the phase angle).

\vspace{-0.2cm}
\subsection{Previous work}
\label{ssec:Previous}

A number of recent papers on NILM present solutions based on deep neural networks. %When examining current work on NILM we can divide solutions roughly into deep-learning based disaggregators and non deep-learning based ones. 
Kelly \textit{et al.} first presented a deep learning-based approach in~\cite{kelly} and since then, a number of other attempts have been made at solving NILM using deep neural networks~\cite{novel,valenti,R-LSTM}. Although such solutions generally achieve good performance, other approaches still remain relevant in NILM. Some successful approaches include Hidden Markov Models (HMM)~\cite{stephen}, as well as the cross entropy method~\cite{CrossEntropy}, and integer  programming~\cite{aidedlinear,mixedlinear}. 

Among the recent NILM approaches using neural networks, convolutional  networks~\cite{valenti} and bi-directional recurrent neural networks~\cite{kelly,R-LSTM} have been proposed. All these approaches are non-causal, requiring future data in order to disaggregate the current sample. In practical applications this means significant delay in disaggregation, effectively preventing real-time use. Our solution is causal, as well as computationally efficient, allowing for true on-line disaggregation.

Other solutions for NILM, such as HMMs, require modeling the problem in a certain manner, limiting the ability to easily adapt the solution to changes in the model or the data. We will demonstrate that performance can be greatly improved by using additional input measurements, a change that would require significant adaptation to~\cite{stephen} and other HMM based NILM solutions. The structure we propose, referred to as WaveNILM, requires no significant alterations whether using one or many input signals. This approach was recently explored in neural networks~\cite{valenti} and in mixed-integer linear programming~\cite{mixedlinear}, demonstrating improved performance when using reactive power along with the standard active power. WaveNILM builds on this, examining the benefits of additional input signals while achieving significantly better disaggregation performance than these earlier works and maintaining causality.

\begin{figure}[tb]
\centering
\centerline{\includegraphics[width=\linewidth,keepaspectratio]{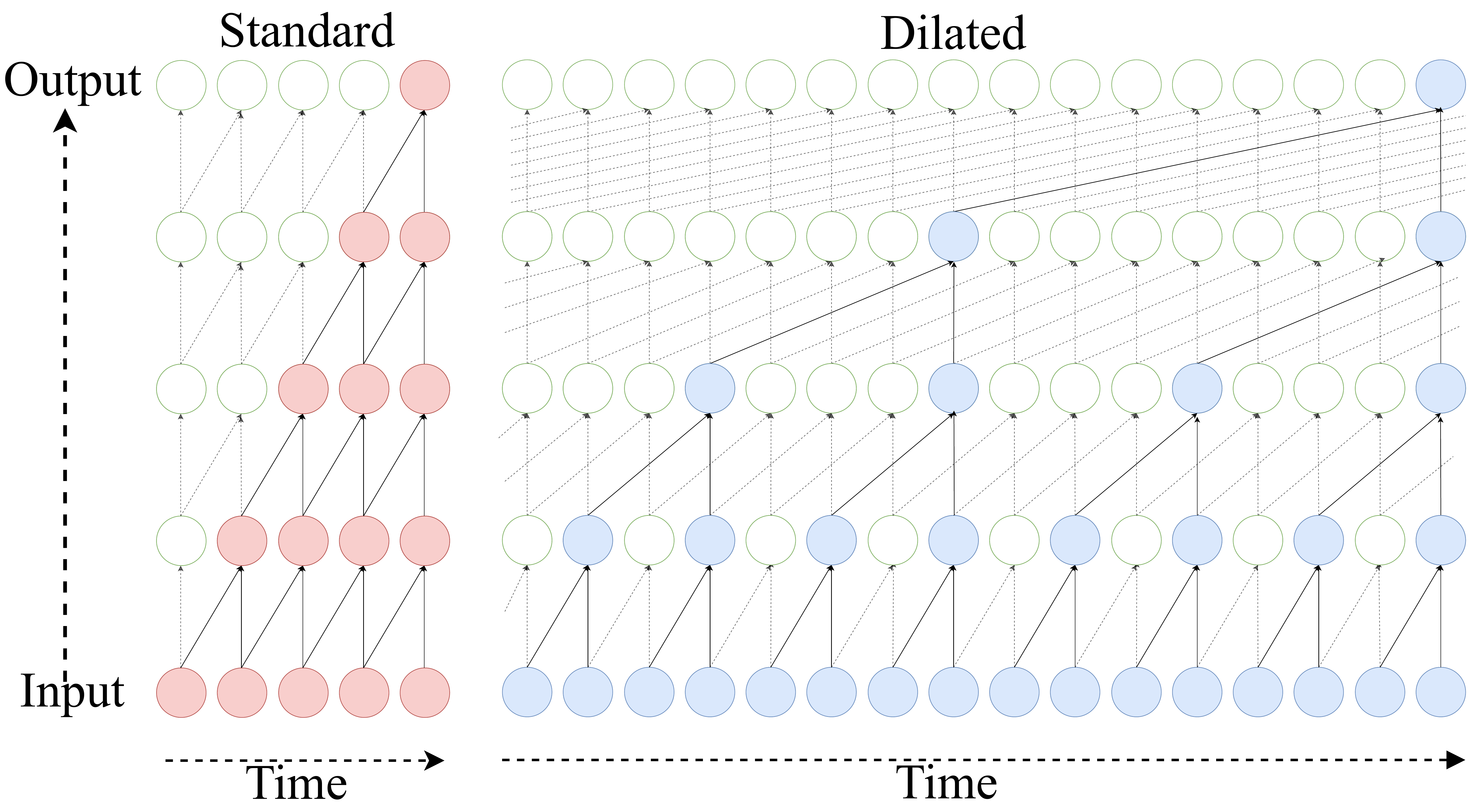}}
\caption{Causal standard (left) and dilated (right) convolution stacks. Both have 4 layers, each with filter length 2. The dilation factor is increased by a factor of 2 with each layer. Colored nodes represent how output is calculated. Choices of colour simply differentiate one network from another and have no other significant meaning.}
\label{fig:dilated}
\vspace{-0.5cm}
\end{figure}

\vspace{-0.3cm}

\section{Proposed Solution}
\label{sec:Solution}
\vspace{-0.2cm}
\subsection{Dilated causal convolutional neural networks}
\label{ssec:ACNN}
Maintaining causality is important in NILM as it allows for disaggregated data to be made available to users in real-time. This is especially true in a dynamically priced power grid, where a user might turn on a high energy appliance at a time where power is expensive. If given a real-time notification, the user may choose to defer the task to a later time, saving money and reducing load on the network at peak time~\cite{ehrhardt2010advanced}.

Causal convolutional neural networks (CNNs) differ from standard CNNs by using only samples from previous time-steps to calculate the current output. A stack of standard causal convolution layers requires long filters or very deep structures in order to achieve sufficiently large receptive fields. One way to address this problem is dilated causal convolution, as introduced in WaveNet~\cite{wavenet}. In a dilated causal convolution with $x[n]$ as the input, dilation factor $M$ and length $N$ with parameters $c_k$, the output $y[n]$ is: 
\begin{equation}
    y[n]=\sum_{k=0}^{N-1}c_k\cdot x[n-M\cdot k].
\end{equation}
Note that receptive field of $y$ is of size $M\cdot(N-1)+1$, even though it only has $N$ parameters. By stacking dilated causal convolutions with increasing dilation factors we can achieve large receptive fields with a limited number of parameters while still maintaining causality, sampling rate, and using all available inputs. Fig.~\ref{fig:dilated} provides a visualization of %the advantages of 
dilated convolution stacks.

\vspace{-0.3cm}

\subsection{Proposed network structure}
\label{ssec:Network}

The basic building block of WaveNILM is a gated version of the dilated causal convolutional layer, also inspired by WaveNet~\cite{wavenet}. Samples from the current and past time steps are used as input to the dilated casual convolutions. Then, the output of each convolutional layer is used as an input to both a sigmoid activation (the relevance estimator or ``gate'') and a rectified linear activation (the regressor). The two activation values are then multiplied, and used as the output of the block. Each block output is then duplicated, one part being used as an input to the next layer, while the other (the ``skip connection'') skips all subsequent convolutions and is used in the final layers of WaveNILM. Each of these layers also contains a dropout of 10\%.
%\textcolor{red}{Is this ``gate'' novel, relative to what was presented in WaveNet~\cite{wavenet}? If so, then it makes more sense to show this in Fig. 1, rather than dilated convolutions (which are available in the WaveNet paper anyway).} \textcolor{blue}{It isn't, they use different activation for the regressor part (a tanh) and they also use residual, but other they also use a gate, that's why initially I hadn't cited wavenet in the section about dilated convolutions, but we can cite in both}

\begin{figure}[tb]
\centering
\centerline{\includegraphics[width=\linewidth,keepaspectratio]{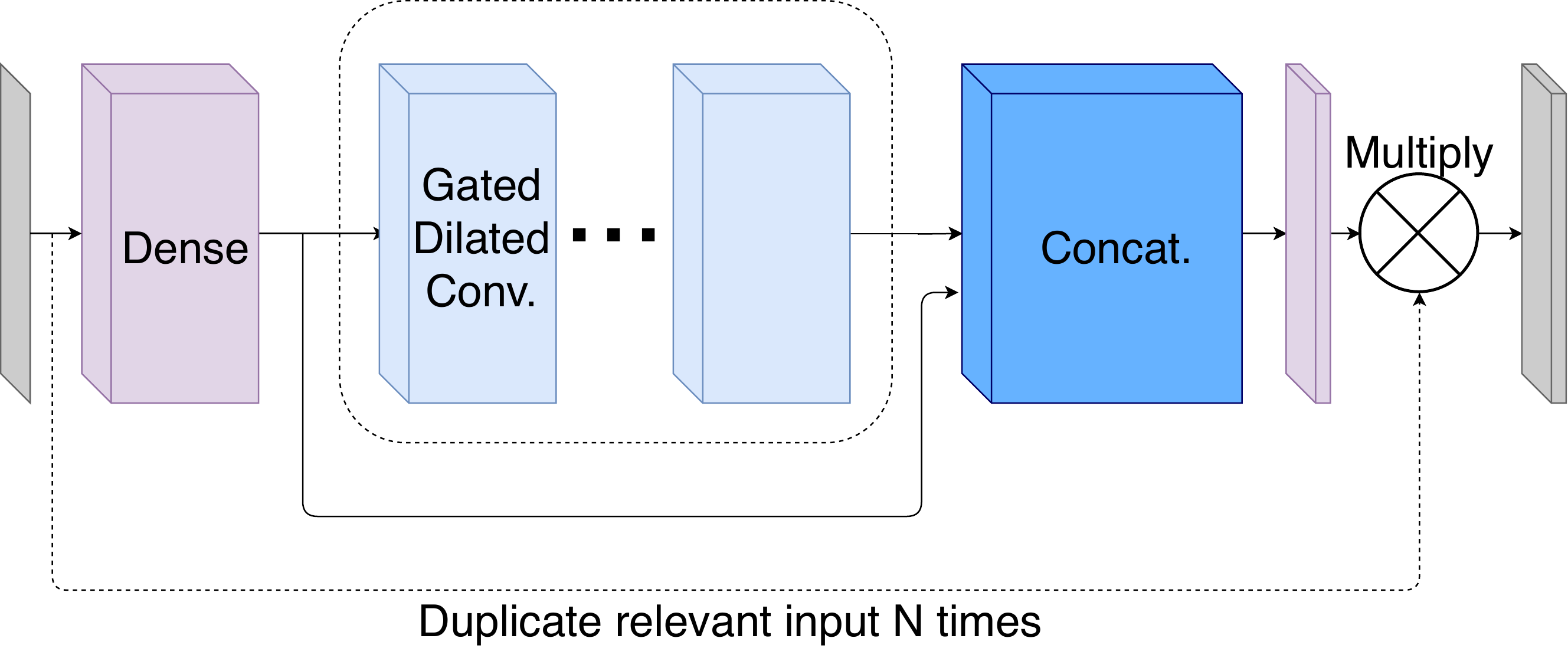}}
\caption{WaveNILM network structure}
\label{fig:WaveNILM}
\vspace{-0.5cm}

\end{figure}

In order to determine the appropriate architecture of WaveNILM, we first examined the size of the receptive field, which relates to the length of meaningful temporal relationships in the data. We found the appropriate length to be 512 samples, which is approximately 8.5 hours with 1-minute sampling. %We then compared layer width (meaning same amount of filters in each building block) of several sizes with varying layer width. While similar performance could be achieved by both option, a varying width structure did so with significantly less parameters. Lastly we considered masked regression, where we create an output mask which is then multiplied by the input to create the final regression. This approach achieved slightly better performance than direct regression.
We then examined various possibilities for the number of layers and the layer size (number of filters) to arrive at the WaveNILM model architecture. In the final configuration (Fig.~\ref{fig:WaveNILM}),the inputs are first fed into a 512-node time-distributed fully connected layer (with no connections between separate time samples), followed by a stack of 9 gated building blocks with 512, 256, 256, 128, 128, 256, 256, 256, 512 filters each. Skip connections from each layer of the stack are then concatenated, followed by another fully connected layer with $\tanh$ activation used as the output mask. 
Because WaveNILM uses multiple inputs, we multiply the mask only by the input relating to the quantity we wish to disaggregate. In total, WaveNILM contains approximately 3,250,000 trainable parameters, though this can vary slightly depending on input and output dimensionality.
 \vspace{-0.3cm}
\section{Experimental Setup}
\label{sec:test}
\vspace{-0.3cm}

\subsection{Data}
\label{ssec:data}
There are many different datasets for NILM, e.g.~\cite{UK-DALE,ampds2,REDD} and more, with varying properties such as sampling frequency, duration, location, etc. We chose to focus on AMPds2~\cite{ampds2}, a low frequency dataset (1 minute sampling), taken from one house in Canada over two years. The measurement includes the aggregate signals and specific data from 20 sub-meters. Each measurement contains current, apparent power, active power, and reactive power. AMPds2 is an extension of a previous dataset known as AMPds~\cite{ampds1}, which contained the same measurements, but only for the first year.

Deferrable loads were defined by~\cite{stephen} as large loads that the user could choose to run at lower levels or at off-peak hours. Successfully disaggregating these appliances is of significant importance to both users, in reducing power cost, and to suppliers, in helping to prevent brownouts by reducing peak time power consumption (known as peak shaving~\cite{peaks}). In AMPds2 these appliances are the HVAC system, the heat pump, the wall oven, the clothes dryer, and the dishwasher. When disaggregating only a subset of all loads, the aggregate signal contains many signals from other loads, as well as the desired signals. We consider these signals from other loads to be the measurement noise (with respect to the desired loads) and consider both noisy and denoised cases. In the noisy case, the inputs are actual aggregate measurements wheareas in the denoised case the inputs are the sum of all target appliance measurements (ground truth signals). On average, a given aggregate reading in AMPds contains about 60\% noise when trying to disaggregate deferrable loads~\cite{stephen}.

Other work using AMPds or AMPds2 uses only denoised scenarios~\cite{mixedlinear} or achieves significantly inferior results~\cite{valenti}. For these reasons, we consider~\cite{stephen}, which is based on a sparse super-state HMM (SSHMM), to be the current state-of-the-art for AMPds and use it for comparison with WaveNILM. At the time of its publication, AMPds2 had not yet been available, meaning~\cite{stephen} was based on AMPds. To ensure a fair comparison with~\cite{stephen}, WaveNILM was trained and tested on the same data and scenarios as~\cite{stephen}, and the results on the full AMPds2 dataset are reported separately. 

We suggest four possible inputs, as collected in AMPds2: current ($I$), active/real power ($P$), reactive power ($Q$), and apparent power ($S$). For each scenario we compare the performance of each input used individually, as well as using a combination of two inputs ($P$ and $Q$), and finally using all four inputs. Household power cost is based, in general, only on active power $P$. For this reason, whenever $P$ was available in the scenario, it was used as the output unless otherwise needed for comparison to previous work. In order to explore the effect of using multiple inputs while still maintaining comparability to previous work, we run various combinations of the parameters of each experiment.

\subsection{Training and testing}
Training was performed on one day samples (1440 time-steps) with a batch size of 50 samples. Data was normalized by the next power of 2 to the maximum of the aggregate data. This can be thought of as scaling the entire meter range to $[0,1]$. Because the receptive field of WaveNILM is 512 samples, disaggregating the present measurement requires the 511 previous samples. To avoid training the network on incomplete data, loss was computed for samples 512-1440. This required creating overlap in the training samples so that all available data was used for both training and testing.

Training was conducted using 90\% of the data and testing on the remaining 10\%. Full disaggregation and deferrable load scenarios were trained for as many as 500 and 300 epochs, respectively, allowing for longer training on the more complex problem.

\subsection{Metrics}
Because waveNILM is a regression network, we focus on the error in disaggregated power, as opposed to the state of the appliances.
A common metric for evaluating disaggregated power, is Estimated Accuracy, defined by Johnson and Willsky~\cite{est.acc} and is one of the well accepted accuracy metrics~\cite{acc.eval}:
\begin{equation}
Est. \space Acc. = 1 - \frac{\sum\limits_{t=1}^{T} \sum\limits_{k=1}^{K} |\widehat{s}_k(t)-s_k(t)|}{ 2\sum\limits_{t=1}^{T} \sum\limits_{k=1}^{K} s_k(t)}, 
\end{equation}
where $\widehat{s}_k(t)$ is the predicted power level of appliance $k$ at time $t$, $s_k(t)$ is the ground truth,  and $T$ and $K$ are the total time and the number of appliances, respectively.
The above expression yields total estimated accuracy; if needed, the summation over $k$ can be removed creating an appliance-specific estimation accuracy~\cite{acc.eval}.

%\begin{table}[!htbp]
%\caption{Full disaggregation scenarios}
%\label{tbl:scen}
%\begin{center}
%\begin{tabular}{|c|c|c|c|c|}

%\hline
%\textbf{In} & \textbf{Out} & \textbf{Loads} &\textbf{Time} & \textbf{Cross Val.}\\ 
%\hline

%I & I & All & 2 Yrs & No\\
%P & P  & All & 2 Yrs & No \\
%Q & Q & All & 2 Yrs & No\\
%S & S & All & 2 Yrs & No\\
%P \& Q & P & All & 2 Yrs & No\\
%All & P & All & 2 Yrs & Yes\\
%All & I & All & 2 Yrs & No\\
%
%
%\hline
%
%\end{tabular}
%\end{center}
%\end{table}
%
%\begin{table}[!htbp]
%\caption{Deferrable loads scenarios}
%\label{tbl:scen2}
%\begin{center}
%\begin{tabular}{|c|c|c|c|c|}

%
%\hline
%
%
%\textbf{In} & \textbf{Out} & \textbf{Loads} &\textbf{Time} & \textbf{Cross Val.}\\ 
%\hline
%
%I & I & Def & 2 Yrs & No\\
%P & P  & Def & 2 Yrs & Yes \\
%Q & Q & Def & 2 Yrs & No\\
%S & S & Def & 2 Yrs & No\\
%P \& Q & P & Def & 2 Yrs & Yes\\
%All & P & Def & 2 Yrs & Yes\\
%All & I & Def & 2 Yrs & Yes\\
%\hline
%\multicolumn{5}{}{}\\
%
%\end{tabular}
%\end{center}
%\end{table}

%\begin{table}[!htbp]
%\caption{Scenarios for comparison with~\cite{stephen}, each repeated as noisy and denoised scenario}
%\label{tbl:scen3}
%\begin{center}
%\begin{tabular}{|c|c|c|c|c|}

%\hline

%\textbf{In} & \textbf{Out} & \textbf{Loads} &\textbf{Time} & \textbf{Cross Val.}\\ 
%\hline
%I & I & Def & 1 Yr & Yes\\
%P & P & Def & 1 Yr & Yes\\
%All & I & Def & 1 Yr & Yes\\
%All & P & Def & 1 Yr & Yes\\
%\hline

%\end{tabular}
%\end{center}
%\end{table}

\vspace{-0.3cm}
\section{Results}
\label{results}
\subsection{Different input signals}
\label{ssec:inputcomp}

We studied the performance of WaveNILM with  a variety of input signals.  When using only one input, the same electrical quantity was used as the output. When using several inputs, we choose active power $P$ or current $I$ as the output. In order to study the significance of input and output signals we compare partial-disaggregation on the deferrable loads with real aggregate input (noisy case), as well as the full 20-load disaggregation.  %As this stage is was an exploratory stage, to find the best configuration for WaveNILM, not all scenarios were 10-fold cross-validated. Those that were are marked by an underline in Table~\ref{tbl:inputs}

%In this experiment, some scenario were not 10-fold cross-validated, those that were however, are marked by an underline in Table~\ref{tbl:inputs}. 

\begin{figure}[tb]
\centering
\centerline{\includegraphics[width=\linewidth,keepaspectratio]{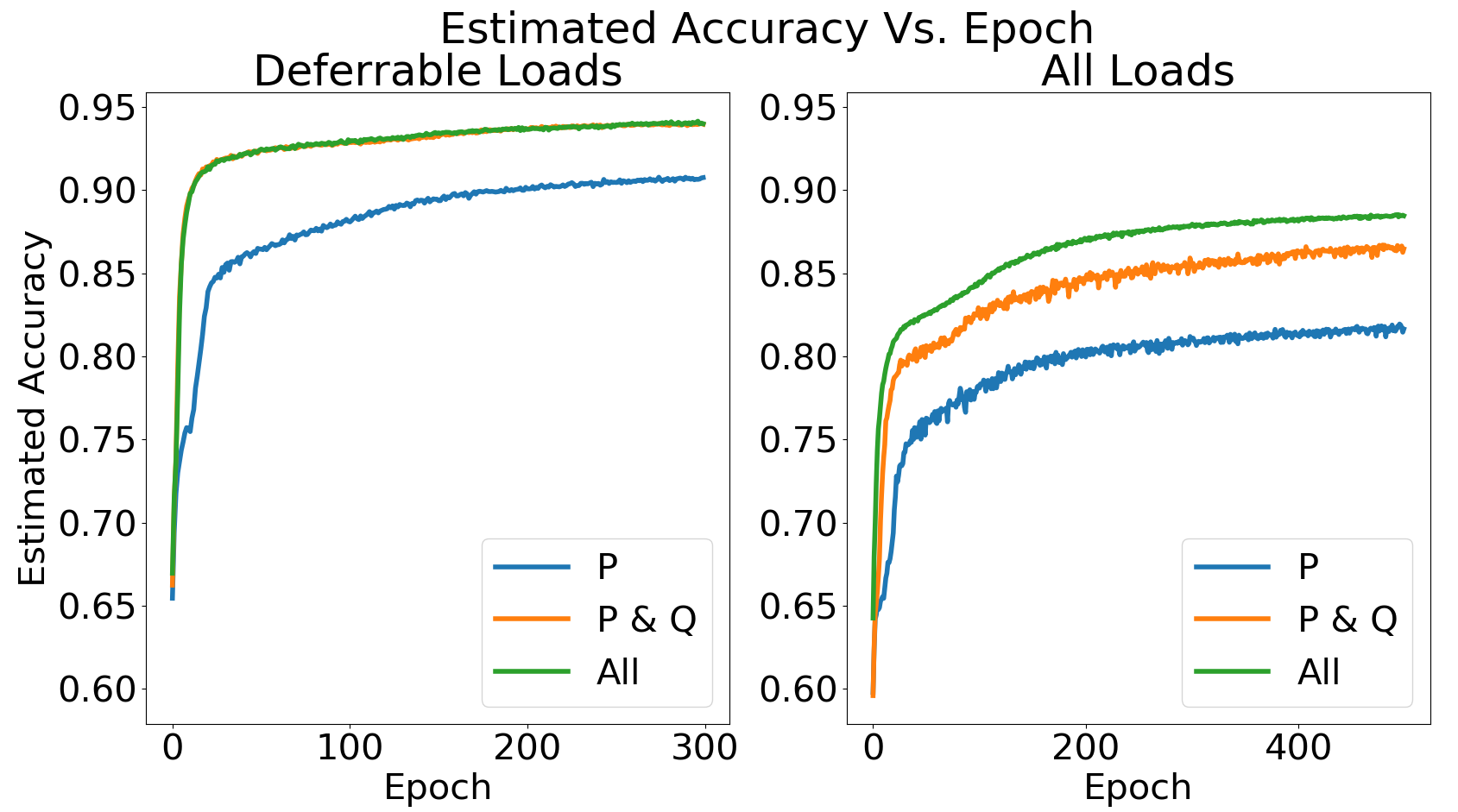}}
\vspace{-0.2cm}
\caption{Convergence speed for single vs. multiple inputs. Note that the on the deferrable loads case (left), the training with P \& Q and training with all inputs produce almost identical curves.}
\label{fig:converge}
\vspace{-0.4cm}
\end{figure}

Examination of the results, as summarized in Table~\ref{tbl:inputs}, shows the significant improvement in performance with the use of multiple inputs. In addition to final performance, convergence time was also greatly improved when using additional inputs, as seen in Fig.~\ref{fig:converge}. In both aspects, the main benefit is achieved when introducing reactive power $Q$ as an input, with subsequent additions providing marginal improvement. Note that $Q$ as a sole input provides excellent results, however since domestic consumers are not charged for it, it's disaggregation alone does not help achieve the goals of NILM as explained in Section~\ref{sec:intro}.

%Note that reactive power as a sole input provides excellent results, however it is normally not available as a reading from the smart meter because domestic consumers are not charged for reactive power. \textcolor{red}{I think our main point shouldn't be it's availability, but rather the fact that it isn't used for billing, we are using it as an input, so we don't want to highlight the fact it's less available than others, though we do mention it later in the conclusions, this is my alternative suggestions:} \textcolor{orange}{(SM) This looks OK to me, because there maybe a time int he future where it is a part of billing for residential customers and the meter firmware is upgraded to send it out.} 

%It is important to note that the case of using reactive power as the only input is an exception to this rule. While this result may prove useful when attempting to classify appliance states, in and of itself, it is lest significant because domestic consumers are not charged for reactive power, which as explained in Section~\ref{ssec:power} eventually flows back to the power grid.

\begin{table}[htbp]
\caption{Noisy case results with various inputs, AMPds2}
\vspace{-0.5cm}
\label{tbl:inputs}
\begin{center}
\begin{tabular}{|c|c|c|}
\hline
\textbf{Input Signal} & \textbf{All loads} & \textbf{Deferrable loads} \\
\hline
$I$ & 85.6\% & 92.0\%\\
$P$ & 82.6\% & 90.9\% \\
$Q$ & \textbf{91.1\%} & 94.4\%\\
$S$ & 86.7\% & 88.9\%\\
\hline
$P$ and $Q$  & 87.5\% & 93.9\%\\
\hline
All 4 (output: $P$) & 88.4\% & 94.2\%\\
All 4 (output: $I$) & \textbf{90.2\%} & \textbf{95.0\%}\\
\hline
\end{tabular}
\end{center}
\vspace{-1.1cm}
\end{table}

\subsection{Comparison with SSHMM}
\label{ssec:comp}
When comparing with SSHMM~\cite{stephen}, only the first year of AMPds2 (i.e., AMPds) was used and the results were all 10-fold cross-validated meaning we repeated each experiment ten times, changing the test set for each iteration, and finally, average the performance over all ten tests.  Both noisy and denoised cases are examined, as explained in Section~\ref{ssec:data}. For each case, the output is the current $I$, as in~\cite{stephen}. The input is either $I$, as in~\cite{stephen}, or all four signals $I,P,Q,S$. %  we present the results of using the all input signals as well as only current or active power as input and either current or active power as an output.
Denoised case results are shown in Table~\ref{tbl:denoised}, with the best result indicated in bold. As seen in the table, the performance on denoised aggregate signals was near perfect.%, even when using a single measurement. 
We consider this problem largely solved. %, and focus on the noisy case.

\begin{table}[htbp]
\caption{Denoised case results on deferrable loads, AMPds}
\vspace{-0.4cm}
\label{tbl:denoised}
\begin{center}
\begin{tabular}{|c|c|c|c|}
\hline
\textbf{Disaggregator} & \textbf{Input} & \textbf{Output} &\textbf{Est. Acc.} \\
\hline
WaveNILM & $I$ & $I$ & 99.0\%\\
%WaveNILM & Active - P & Active - P & 98.9\% \\
%\hline
WaveNILM & $I, P, Q, S$ & $I$ &\textbf{99.1}\%\\
%WaveNILM & All 4 Inputs & Active - P & 98.9\%\\
\hline
SSHMM~\cite{stephen} & $I$ & $I$ & 99.0\% \\
\hline
\end{tabular}
\end{center}
\vspace{-0.8cm}
\end{table}

Next we consider the noisy case, where the actual aggregate measurement was used as an input. Again we compare the same configurations of WaveNILM as in the denoised case, and all experiments were 10-fold cross-validated. As seen in Table~\ref{tbl:noisy}, when compared with SSHMM~\cite{stephen}, WaveNILM performs slightly worse when using only $I$ as an input, and slightly better when using all four inputs. Run times are also improved, with WaveNILM completing disaggregation of one sample in under 150$\mu s$ for all configurations, where the SSHMM required approximately 4.55$ms$ (though both are appropriate for real-time disaggregation). More importantly, the run time (and storage space) of WaveNILM increases only marginally when increasing the number of loads, the difference between the simplest configuration (1 input, 5 loads) and most complicated (4 inputs, all 20 loads) being under 20$\mu s$.

\begin{table}[htbp]
\caption{Noisy case results on deferrable loads, AMPds}
\vspace{-0.4cm}
\label{tbl:noisy}
\begin{center}
\begin{tabular}{|c|c|c|c|}
\hline
\textbf{Disaggregator} & \textbf{Input} & \textbf{Output} &\textbf{Est. Acc.} \\
\hline
%WaveNILM & Active - P & Active - P  & 90.3\%\\
WaveNILM & $I$ & $I$  & 92.3\%\\
%\hline
%WaveNILM & All 4 & Active - P  & 93.9\%\\
WaveNILM & $I, P, Q, S$  & $I$  & \textbf{94.7}\%\\

\hline
SSHMM~\cite{stephen} & $I$ & $I$ & 94.0 \%\\
\hline
\end{tabular}
\end{center}
\vspace{-1.1cm}
\end{table}

\section{Conclusions}
\label{sec:conclusion}
We presented WaveNILM, a flexible and causal CNN for load disaggregation. One of its greatest advantages over existing NILM solutions is the ability to easily add various inputs to help improve disaggregation. WaveNILM requires only 512 new parameters for each new input signal, and only 3072 parameters for each added output (load). This is particularly important when exploring additional external measurements that have been shown to be useful for disaggregation such as weather data~\cite{mmsp18}, or time of day~\cite{dinesh}.
%We have shown that WaveNILM performs better than state-of-the-art on the same data, with a large part of the improvement resulting from using all electrical properties of the measured aggregate signal improves performance. In addition, WaveNILM allows for easy adaptation to change in model, requiring only 512 new parameters for each new input signal, and only 3072 parameters for each added load. This is particularly important when exploring the incorporation of external measurements that have been shown to be useful for disaggregation such as weather data~\cite{mmsp18}, or time of day~\cite{dinesh}.

Reactive power has been shown to be of particular benefit to NILM, providing most of the performance benefit, with apparent power and current only adding marginal improvement. As far as data collection is concerned, we believe the results presented here should encourage the recording of as many electrical signals as possible, with active and reactive power being the most significant. Reactive power is not currently reported by most smart meters, yet it can be calculated within the meter. This means that reactive power is, in principle, an available measurement, and could be more accessible in future generations of smart meters.
%if WaveNILM can run on the meter, reactive power should be available for use. We believe this is possible considering the relative simplicity and low sampling-rate data used for WaveNILM. %In the past, we have shown~\cite{mmsp18} that a similarly complex neural network can be implemented on a Raspberry Pi and still run faster than needed for real-time NILM. 

During the review of this paper, another model under the name of ``WaveNILM'' was presented in~\cite{martins2018wave}. While both our model and the one in~\cite{martins2018wave} are derived from WaveNet~\cite{wavenet}, they are quite distinct and intended for different applications:~\cite{martins2018wave} for industrial loads, ours for residential loads.
Finally, as of June $18^{th}$ 2019, the code for WaveNILM is available at https://github.com/picagrad/WaveNILM .

\normalem

\bibliographystyle{IEEEbib.bst}
\bibliography{strings,refs}

\end{document}